\documentclass{sig-alternate-10pt}

\usepackage{graphicx}
\usepackage{balance}
\usepackage{amssymb, amsmath}
\usepackage{multirow, rotating, wasysym, url}
\usepackage[tight]{subfigure}
\usepackage[ruled,linesnumbered,vlined]{algorithm2e}
\usepackage{hyperref} %reference hyper links.
\usepackage{color}
\usepackage{float}
\usepackage{caption}
\usepackage{enumitem}
\graphicspath{{./DrawingFigures/},{./EvaluateFigures/}}

	\newcommand{\presec}{\vspace{-0.07in}}
	\newcommand{\postsec}{\vspace{-0.05in}}
	\newcommand{\presub}{\vspace{-0.07in}}
	\newcommand{\postsub}{\vspace{-0.05in}}

	\newcommand{\postfig}{\vspace{-0.05in}}

		\mathchardef\Gamma="0100 \mathchardef\Delta="0101
\mathchardef\Theta="0102 \mathchardef\Lambda="0103
\mathchardef\Xi="0104 \mathchardef\Pi="0105
\mathchardef\Sigma="0106 \mathchardef\Upsilon="0107
\mathchardef\Phi="0108 \mathchardef\Psi="0109
\mathchardef\Omega="010A

\newcommand{\outline}[1]{}%{\textbf{#1}}

\usepackage{cite}
\usepackage{xspace}
\usepackage{url}
\usepackage{graphicx}
\usepackage{latexsym}
\usepackage{amssymb}
\usepackage{amsfonts}
\usepackage{psfrag}
\usepackage{wrapfig}
\usepackage{comment}
\usepackage{alltt}
\usepackage{color}

\newcommand{\eg}{\emph{e.g.}\xspace}
\newcommand{\etc}{\emph{etc.}\xspace}

\newcommand{\Comment}[1]{}

	\newcommand{\para}{\noindent\textbf}
	
	\newcommand{\figwidthdraw}{0.4\textwidth}

\usepackage[normalem]{ulem}

\title{FISF: Better User Experience using Smaller Bandwidth for Panoramic Virtual Reality Video}
\author{Lun Wang$^\text{\dag}$, Damai Dai$^\text{\dag}$, Jie Jiang$^\text{\dag}$, Tong Yang$^\text{\dag}$, Xiaoke Jiang$^*$, Zekun Cai$^\text{\dag}$, Yang Li$^*$, Xiaoming Li$^\text{\dag}$\\
		$^\text{\dag}$ Peking University, $^*$ Kandao Technology Co., Ltd.}

\begin{document}
\maketitle
\sloppy
	\begin{abstract}

The panoramic video is widely used to build virtual reality (VR) and is expected to be one of the next generation Killer-Apps.
Transmitting panoramic VR videos is a challenging task because of two problems: 
1) panoramic VR videos are typically much larger than normal videos but they need to be transmitted with limited bandwidth in mobile networks.
2) high-resolution and fluent views should be provided to guarantee a superior user experience and avoid side-effects such as dizziness and nausea.
To address these two problems, we propose a novel interactive streaming technology, namely Focus-based Interactive Streaming Framework (FISF).
FISF consists of three parts: 1) we use the classic clustering algorithm DBSCAN to analyze real user data for Video Focus Detection (VFD); 2) we propose a Focus-based Interactive Streaming Technology (FIST), including a static version and a dynamic version; 3) we propose two optimization methods: focus merging and prefetch strategy.
Experimental results show that FISF significantly outperforms the state-of-the-art.
The paper is submitted to Sigcomm 2017, VR/AR Network on 31 Mar 2017 at 10:44:04am EDT. 

\end{abstract}
	\vspace{-0.05in}
\presec
\section{Introduction}
\postsec

\presub
\subsection{Background and Motivation} \postsub
The panoramic video is widely used to build virtual reality (VR) and is expected to be one of the next generation Killer-Apps. %\cite{Sigcomm}
It is widely used in online multimedia \cite{googlestreet}, video surveillance \cite{videosurveillance} and robotics \cite{robotics1} applications because of its good interactivity.

\begin{figure}[h]
\centering
\postfig
\includegraphics[width=1\linewidth]{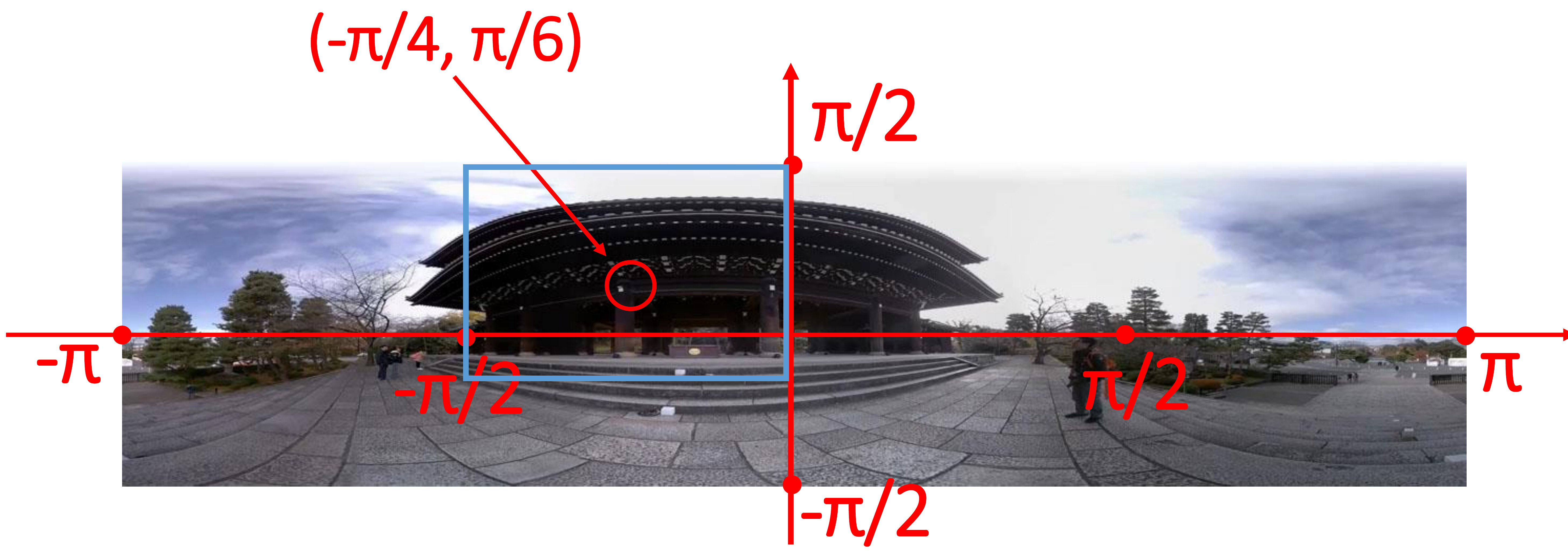}
\caption{Illustration of Coordinate.}
\label{draw:coordinate}
\postfig
\end{figure}

As shown in Figure \ref{draw:coordinate},
a panoramic VR video is typically a two-dimensional rectangular video. Video players will map it onto a mesh (typically, sphere or skybox \cite{skybox}), and render it on users' screen like helmet-mounted devices (HMD) or mobile phones. 
When watching a panoramic VR video, the users can navigate the scenes interactively by changing their viewpoints and viewing directions. 
One point in a video is described by a quadruple: $(t, x, y, z)$. $t$ is the timing of the video, $(x, y)$ are the spatial coordinates, and $z$ is the users' viewing directions.

Transmitting panoramic VR videos is a challenging task because of two problems:
1) panoramic VR videos are typically much larger than normal videos (fluent transmission requires 10+Mbps) but they need to be transmitted with limited bandwidth in mobile networks. 
2) high-resolution views should be provided to guarantee a superior user experience and avoid side-effects such as dizziness and nausea.

\vspace{-0.04in}
\presub
\subsection{Limitations of Prior Art} \postsub

The naive solution directly transmits $360^\circ\times180^\circ$ high-resolution panoramic VR videos, which consumes much bandwidth and causes \texttt{lag phases}, where lag phase means that the users' screens keep unchanged or vague for seconds.
Facebook provides a solution called interactive streaming technology \cite{Facebook}.
It produces 32 copies of videos, each of which contains only a \texttt{fixed} high-resolution area (\eg, $90^\circ\times120^\circ$ scene, see the blue rectangle in Figure \ref{draw:coordinate}), while the other areas are low-resolution.
It chooses the most appropriate copy and transmits it to users based on the users' current viewpoints. In this way, the bandwidth for transmission is significantly reduced.  
However, when the users' viewpoints change, a new copy needs to be transmitted. Transmission latency inevitably incurs lag phases on the users' screens. The lag phases, caused by the transmission latency (typically for seconds), will significantly degrade the user experience because the users can only watch low-resolution panoramic VR videos during lag phases.
Because of the method's large influence and practicability, we consider it as the state-of-the-art.
Another solution is proposed by Ochi Daisuke and several other researchers \cite{basicflowswitching}, which is similar to Facebook's solution. 
There are several other cutting-edge researches aiming at addressing the bandwidth and user experience issues, such as object-based transcoding \cite{ObjectBasedEncoding}, perception-based scheduling \cite{PerceptionBasedSchedualing}, and active video annotations \cite{MultiplePerspectives, VideoAnnotation}. They are based on different technologies, such as object detection, annotation, \etc Our proposed solution is different from these researches because it is based on analysis of real-world user-watching traces.

%The \texttt{design goal} of this paper is to reduce the total time of lag phases and improve user experience with low bandwidth.

\presub
\subsection{Our Proposed Solutions} \postsub

In this paper, we propose a \texttt{focus-based interactive streaming framework} (FISF). FISF consists of a \texttt{video focus detection} (VFD) algorithm based on user data analysis, a static and a dynamic \texttt{focus-based interactive streaming technologies} (FIST), and two further optimizations: focus merging and prefetch strategy. FISF achieves a much smaller number of lag phases, and makes users enjoy high-resolution views with low bandwidth.

%provides more high-resolution frames in users' screen.

\begin{figure}[h]
\centering
\postfig
\includegraphics[width=1\linewidth]{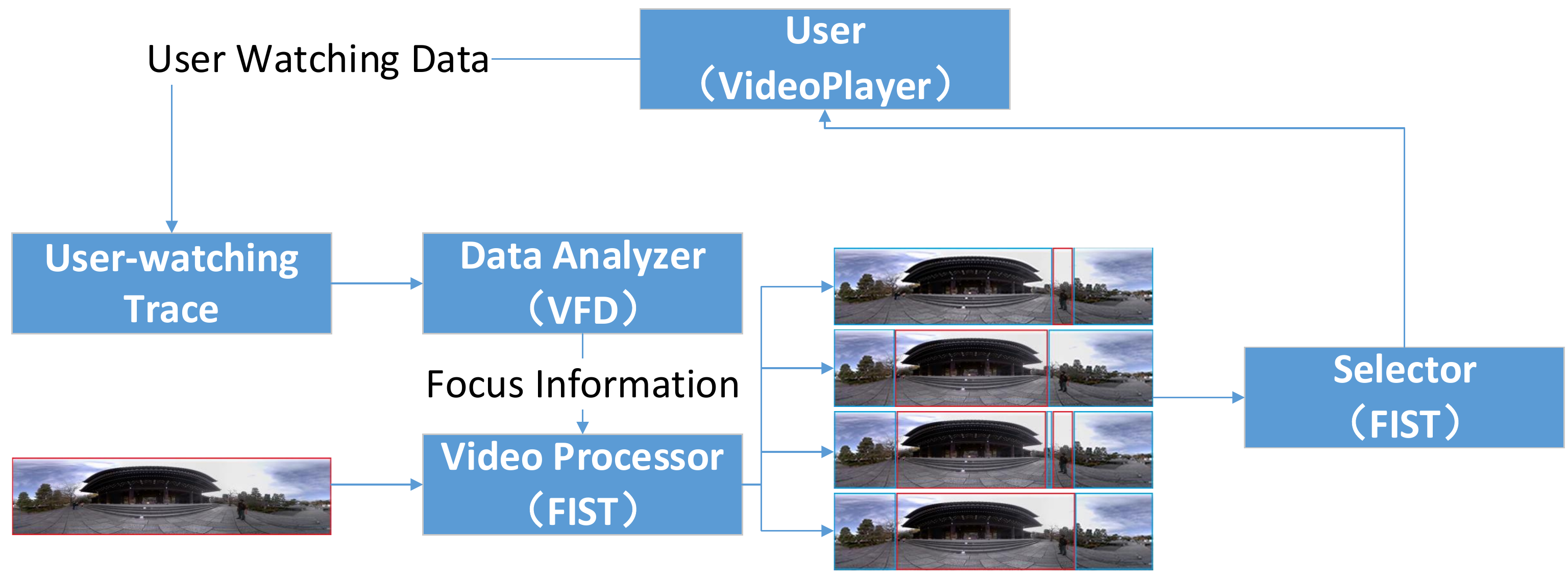}
\caption{Hierachy of FISF.}
\label{draw:hier}
\postfig
\end{figure}

FISF is based on our key observation from the tests and analysis of real panoramic videos: when the users watch a panoramic video, there are some viewpoints more likely to be watched for a long time, namely \texttt{focuses}.
The framework of FISF is shown in Figure \ref{draw:hier}.
First, we propose a VFD algorithm to detect video focuses, where VFD leverages the well-known DBSCAN \cite{dbscan} algorithm. Second, similar to the state-of-the-art \cite{Facebook}, we also produce multiple video copies, and each copy also contains a small area of high-resolution video. 
In the state-of-the-art solution, the high-resolution area is irrelevant to the video focuses. While in our solution, the high-resolution area of each copy contains one or more video focuses.
%The state-of-the-art \cite{Facebook} only saves bandwidth. In contrast, FISF reduces total lag phase time, and also saves more bandwidth. 
The state-of-the-art \cite{Facebook} saves bandwidth at a high cost of large switching number and long lag phase time. In contrast, FISF reduces switching number and lag phase time, and even saves more bandwidth.%bandwidth-friendly transmission of interactive streaming technology \cite{basicflowswitching}, 
%FISF is based on the empirical observation that when users watch a panoramic video, there are some viewpoints more likely to be watched for a long time, namely \texttt{focuses}. If the videos can be divided into tiles based on the focus distribution, 

\presub
\subsection{Key Contributions} \postsub
%We make following four key contributions in this paper.
%
\begin{itemize}[leftmargin=1\parindent]
\vspace{-0.09in}
\item We propose the idea of detecting video focuses by analyzing user data, and use it to improve panoramic VR video transmission. We also propose a framework to augment video transmission with video focus detection. 

\vspace{-0.09in}
\item We present a concrete algorithm for data-based video focus detection, two versions of focus-based interactive streaming technologies: a static version and a dynamic version, and two further optimizations.

\vspace{-0.09in}
\item
We simulate our framework and perform extensive experiments using real user-watching traces to evaluate the improvement in terms of user experience and bandwidth.

\end{itemize}

	\presec
\section{Methodology}
\postsec
\label{Methodology}

In this section, we present the three parts of FISF.
First, we present a video focus detection method, called \texttt{video focus detection based on user data analysis} (VFD), which uses DBSCAN clustering algorithm \cite{dbscan}. Second, 
to improve the user experience and save bandwidth for VR videos, 
we propose an algorithm, namely  \texttt{Focus-based Interactive Streaming Technology} (FIST) with a static version and a dynamic version. Third, we propose further optimization methods, including focus merging and prefetch strategy.

\presub
\subsection{Part I: VFD}
\label{VFD}
\postsub

In this subsection, we present the first part of FISF, namely VFD. It serves to detect the focuses of videos, and provides the focus information, so as to help the video processor produce different copies of videos. 
As there is no algorithm to detect VR video focuses based on user data, we tried several classic clustering algorithms, and (DBSCAN) \cite{dbscan} exhibits the best performance. According to experimental results on real user data, the focuses detected by VFD highly conform with empirical results and serve well for FIST. 
%To detect focuses, we propose to use the classic clustering algorithm called \texttt{Density-Based Spatial Clustering of Applications with Noise} (DBSCAN) \cite{dbscan}.
%The core technique of VFD is a clustering algorithm called \texttt{Density-Based Spatial Clustering of Applications with Noise} (DBSCAN) \cite{dbscan}. 

\begin{figure}[h]
\centering
\postfig
\includegraphics[width=1\linewidth]{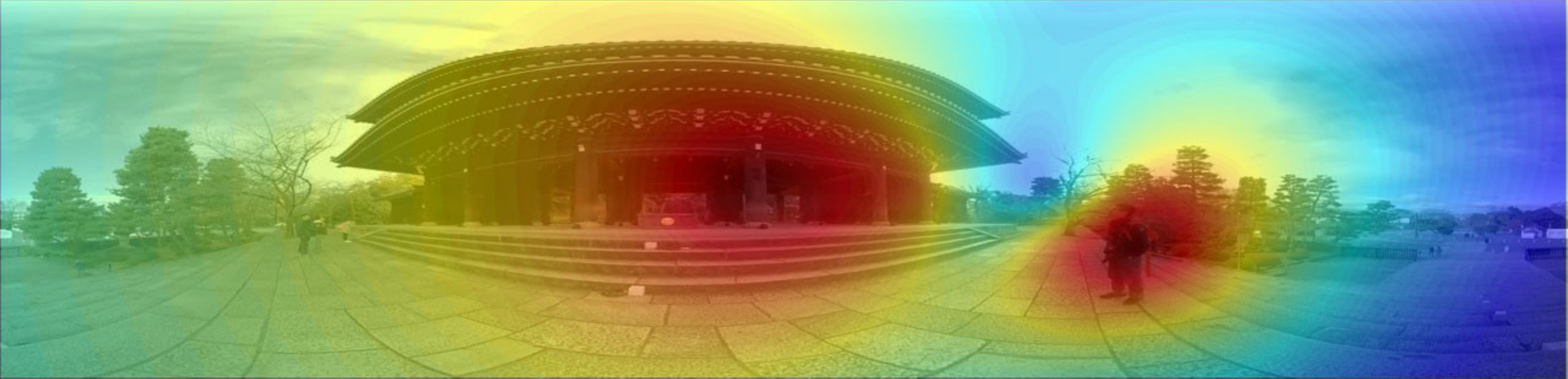}
\caption{Probability Distribution of Attention. Red area means high attention probability. Blue area means low attention probability.}
\label{draw:APD}
\postfig
\end{figure}

\para{Rationale:} Our key observation is that users tend to focus on only some specific points, and ignore other parts when watching a video, especially panoramic VR video, because only part of the video can be seen. The intuition is further confirmed by a simple analysis on real user data. For example, when users look at the picture shown below, the empirical probability distribution of attention is shown in Figure \ref{draw:APD}. Those maximum points with the highest probability to be focused on are focuses. There are two approaches to achieving focus detection: 1) based on computing graphics and 2) based on data analysis. In this paper, we choose the second solution because of two reasons: 1) the second solution is more accurate than the first one because it is based on real user-watching traces; 2) the second solution is more time-efficient in terms of algorithm complexity.

\para{DBSCAN:} VFD chooses DBSCAN to detect focuses. The DBSCAN algorithm views clusters as areas of high density separated by areas of low density, where density is defined by two parameters \texttt{min-samples} and \texttt{eps} \cite{dbscan}. The algorithm examines every sample and find its neighbors (which means samples within a distance of eps). If the number of neighbors is larger than min-samples, we say the area near the determined sample is dense and call the sample \texttt{core sample}. If a sample is not a core sample but it is a neighbor of a core sample, we still put it in the cluster. However, if a non-core sample does not have any core sample neighbors, it is not part of any cluster. 
The key challenge using DBSCAN for focus detection is the selection for parameters. We address this problem using a validation part in VFD. 

\begin{figure}[h]
\centering
\postfig
\includegraphics[width=1\linewidth]{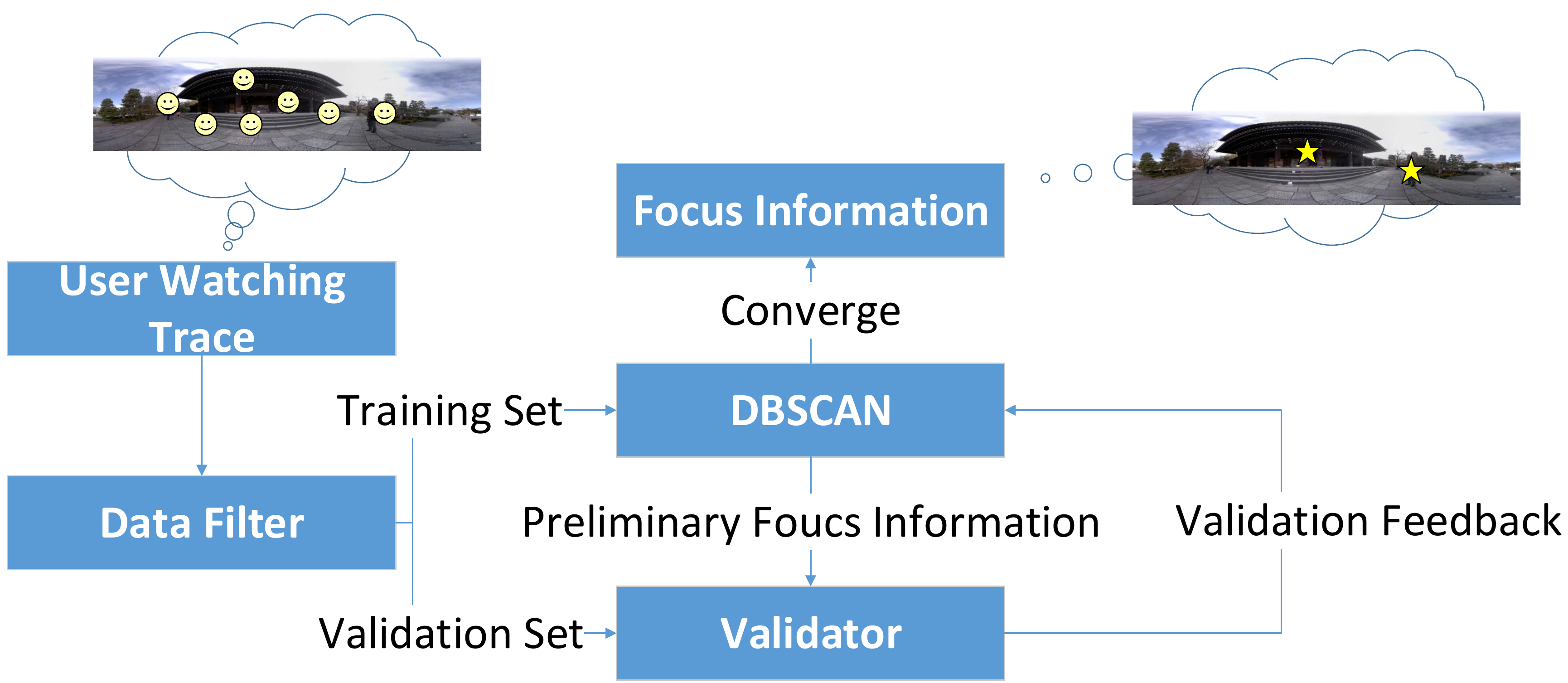}
\caption{Workflow of VFD.}
\label{draw:VFD}
\postfig
\end{figure}

\vspace{0.09in}
\para{VFD:} Figure \ref{draw:VFD} shows the flow chart of VFD and Table \ref{table:watchingrecorddata} shows the user-watching trace format. The VFD algorithm is composed of the following three steps. 1) Data filtering. Data filtering preprocesses the user-watching traces and filters out the ``\texttt{dirty data}'' such as a long-time lag phase without any interactive behavior. Then the ``\texttt{clean data}'' are divided into two sets: a training set and a validation set. 
2) Clustering. DBSCAN is applied on the training set with preset parameters to provide ``preliminary ''. 3) Validation. These focuses, combined with the validation set, are used to simulate the real user behaviors. This will produce feedback to the DBSCAN algorithm and new parameters will be chosen according to the feedback. This procedure will stop until convergence or it reaches the preset upper bound of iterations.

\begin{table}[htbp]
\begin{tabular}{|c|c|c|c|c|}
\hline
\textbf{ID}&\textbf{Timing}&\textbf{x}&\textbf{y}&\textbf{z}\\
\hline
$101$&0.329&-1.542456&-0.2082523&0.2079071\\
\hline
$101$&1.328&-1.54239&-0.2015937&0.2011556\\
\hline
$102$&0.045&-1.495437&-0.02360264&1.607887\\
\hline
$101$&2.336&-1.541883&-0.198082&0.1975058\\
\hline
$\cdots$&$\cdots$&$\cdots$&$\cdots$&$\cdots$\\
\hline
\end{tabular} \postfig
\centering\caption{Watching Record Data Table.}
\label{table:watchingrecorddata}
\end{table}

\begin{figure}[h]
\centering
\postfig
\includegraphics[width=1\linewidth]{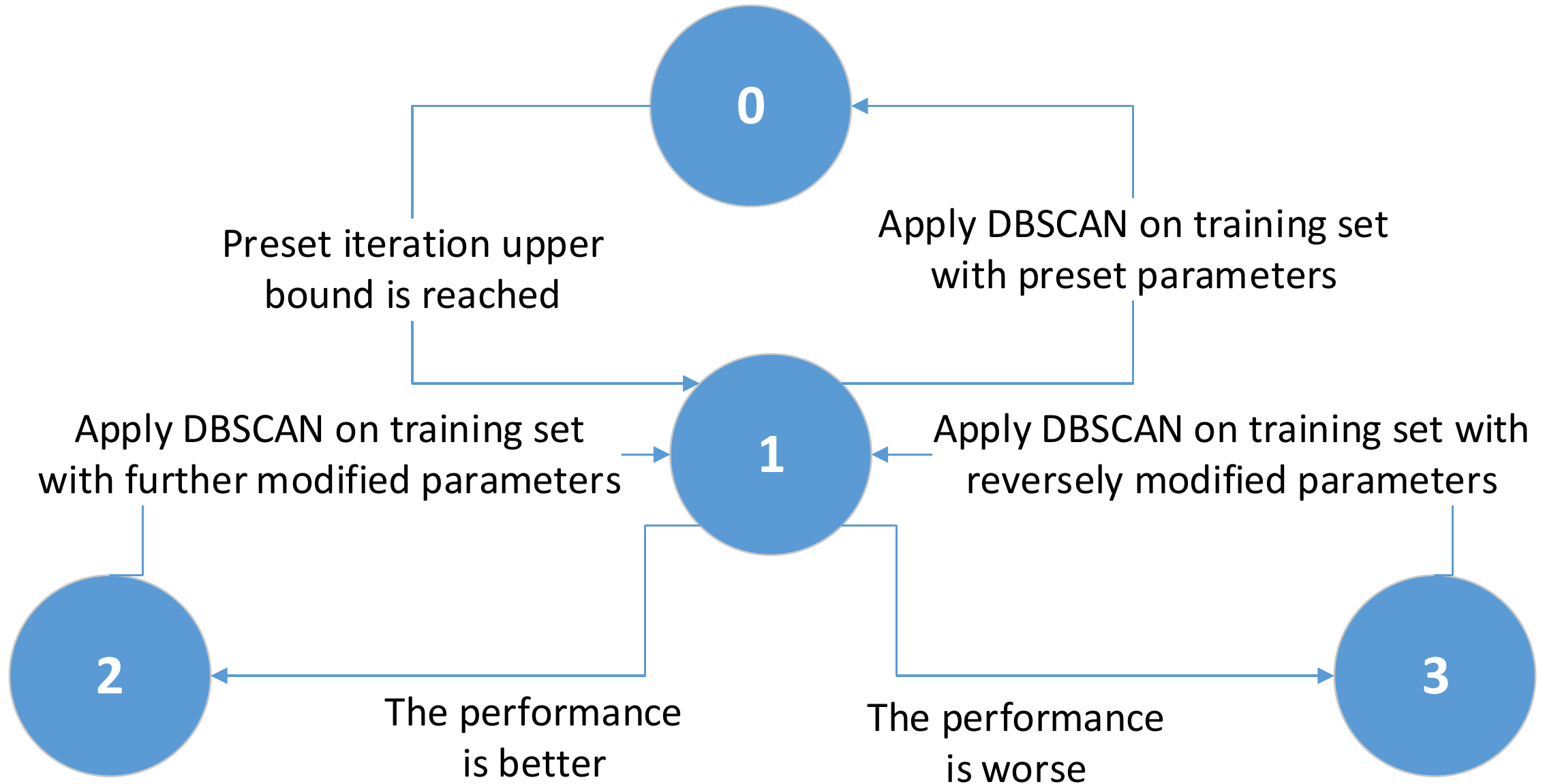}
\caption{FSM of VFD.}
\label{draw:FSM}
\postfig
\end{figure}

The finite state machine (FSM) of DBSCAN and validation is shown in Figure \ref{draw:FSM}. The initial parameters for DBSCAN are preset, so focuses detected are likely to be not accurate, and thus have a poor performance in saving bandwidth and improving the user experience.
To address this issue, we set a validation part to verify the performance by simulating real user behavior using validation set. If the performance is better, we change the parameters in the same direction with a fixed step length (if this is the first time of validation, choose the direction randomly). Otherwise, we change the parameters in the reverse direction with a reduced step length. The procedure stops either when it converges or when the number of iterations reaches the preset upper bound. In section \ref{Experiment}, we carry out an experiment to decide the appropriate preset parameters (see Figure \ref{eva:fnp}).

\presub
\subsection{Part II: FIST}
\postsub

In this subsection, we propose two versions of FIST. We first introduce the basic framework for FIST. Then we introduce two versions of FIST: a static version and a dynamic version, suitable for different videos.

\begin{figure}[h]
\centering
\postfig
\includegraphics[width=1\linewidth]{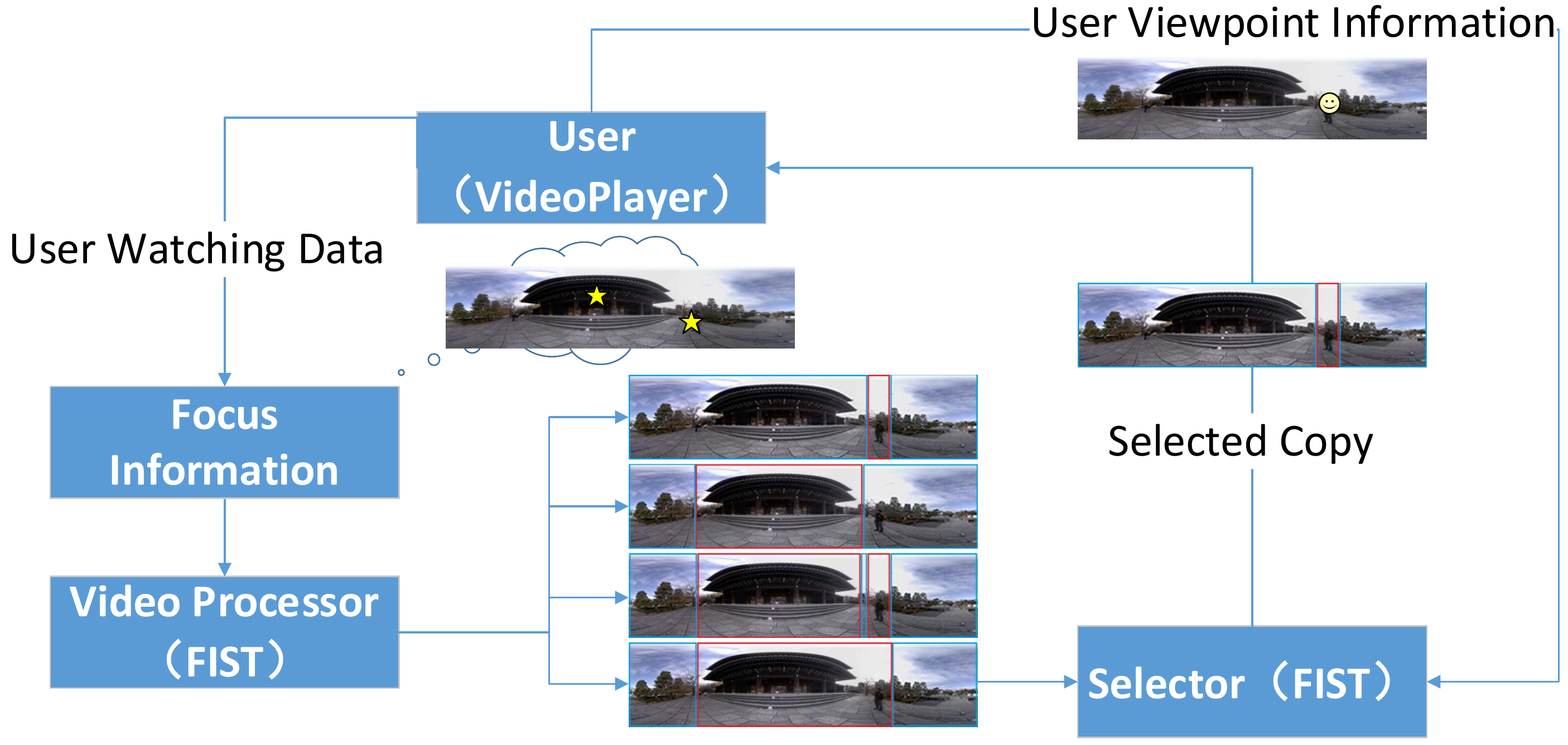}
\caption{Framework of FIST.}
\label{draw:FIST}
\postfig
\end{figure}

\vspace{0.09in}
\para{Basic FIST Framework:} The basic framework of FIST is shown in Figure \ref{draw:FIST}. FIST consists of two main parts: 1) A video processor used to produce different copies of videos; 2) A \texttt{selector} to choose which copy to transmit according to the current user viewpoint. Focus information from VFD is passed to the video processor, and several copies are produced. Each copy, namely \texttt{fcopy} has a high-resolution area covering one or more focuses while other parts are low-resolution. When users watch videos, watching devices like helmet-mounted devices or mobiles phones will detect the users' viewpoints and report them to the selector. The selector will choose the copy with the corresponding video to transmit. Note that the producer also produces four copies, namely \texttt{bcopy}, each with a $90^\circ$ high resolution area and together covering the whole video. When the users' viewpoints are out of any focus, the selector will transmit one of these four copies according to the viewpoints.

This method may introduce extra time and space consumption, but they are both ignorable compared to the bandwidth bonus. First, the preprocessing needs to be done only once. Second, the low-resolution area of a video consumes much smaller memory space than the high-resolution area, so the copies will just consume a little extra memory than a $360^\circ\times180^\circ$ high-resolution video. Third, storage is cheap for panoramic VR video provider so memory usage is also not a problem.

\vspace{0.05in}
\para{Static FIST:} Static FIST uses static focus information to produce video copies. To provide focus information, VFD ignores the time dimension ($t$), and applies DBSCAN on $x$ and $y$ dimension, thus the focuses only contain spatial information. 
The strategy of selecting which copy to transmit is shown in Algorithm \ref{alg:selection}.

\vspace{0.05in}
\para{Dynamic FIST:} Dynamic FIST is based on the key observation that focuses move in a predictable pattern in many videos. For example, in a broadcasting video for a basketball game, the focus is likely to follow the ball. If we apply static FIST to this video, it will switch the copies transmitted back and forth, leading to frequent lag phases. Dynamic FIST addresses this problem by applying DBSCAN on $x$, $y$, and $t$ dimensions. Therefore, the focuses contain time information. When we preprocess the videos to produce copies of the original videos, the copy covering dynamic focuses will have a moving high-resolution area. 
Note that the implementation of Dynamic FIST is still our undergoing work. The selection algorithm for dynamic FIST is shown in algorithm \ref{alg:selection}.

\begin{algorithm}[h]
	\caption{Copy Selection for FIST}
	\label{alg:selection}
	\KwIn{~~User viewpoint: $V_u$\\~~~~~~~~~~~~~Current copy being transmitted: $C_c$}
	\KwOut{Selected Copy}
	
	\If{$V_u$ in the high-resolution area of $C_c$}
	{
		Keep transmitting $C_c$\\
		goto \textbf{Done}\\
	}
	\Else
	{
		$\#$ Dynamic version\\
		$\mathcal{S}_n$ = find dynamic fcopies containing $V_u$\\
		$\#$ Static version\\
		$\mathcal{S}_n$ = find static fcopies containing $V_u$\\
		\If{$|\mathcal{S}_n|$==0}
		{
			$C_n$ = find bcopies containing $V_u$\\
		}
		\Else
		{
			$C_n$ = randomly choose a fcopy from $\mathcal{S}_n$ 
		}
		Switch to transmit $C_n$
	}
	\textbf{Done}
\end{algorithm}

\begin{comment}
\begin{algorithm}[h]
	\caption{Copy Selection for FIST}
	\label{alg:selection}
	\KwIn{User viewpoint: $V_u$\\ Current copy being transmitted: $C_c$}
	\KwOut{Selected Copy}
	
	$\#$ Note that the branch is only active in dynamic FIST\\
	\If{$V_u$ locates in the high-resolution area of a dynamic fcopy $C_d$}
	{
		$\#$ Note that $C_d$ may be the same as $C_c$\\
		Switch to transmit $C_d$
	}
	$\#$ Note that the branch is only active in static FIST\\
	\ElseIf{$V_u$ locates in the high-resolution area of a static fcopy $C_s$}
	{
		$\#$Note that $C_s$ may be the same as $C_c$\\
		Switch to transmit $C_s$
	}
	\ElseIf{$V_u$ locates in the high-resolution area of $C_c$}
	{
		Keep transmitting $C_c$
	}
	\Else{
		Find the bcopy $C_b$ covering $V_u$\\
		Switch to transmit $C_b$
	}
\end{algorithm}
\end{comment}

\para{Static Version vs. Dynamic Version:} Note that these two versions adapt to different situations. When a video has static focuses, static version will definitely have better performance due to more accurate focuses. However, when the focus moves, dynamic version will be better because in static version, the server will need to switch copies frequently and introduce many lag phases. 

\presub
\subsection{Part III: Undergoing Work}
\postsub

In this section, we propose two optimization approaches: focus merging and prefetch strategy.

\begin{figure}[h]
\centering
\postfig
\includegraphics[width=1\linewidth]{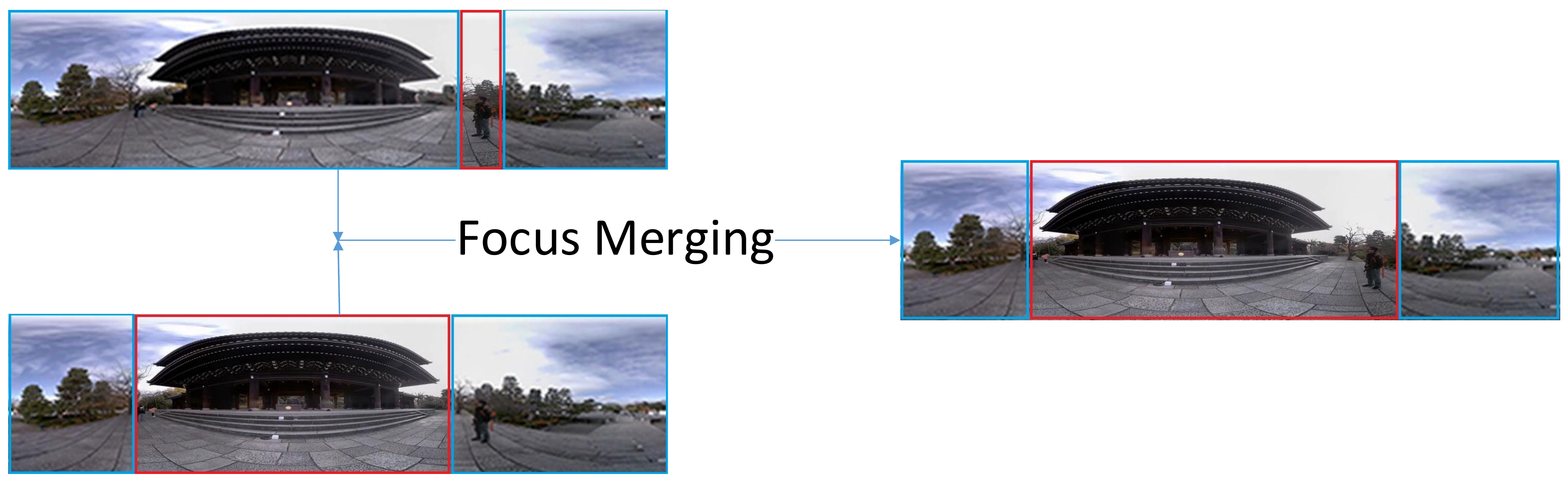}
\caption{A Focus Merging Example: Red part means high-resolution area, blue the opposite.}
\label{draw:focusmerging}
\postfig
\end{figure}

\vspace{0.05in}
\para{Focus Merging:} We observe that users sometimes change frequently between two near focuses. Although one focus may be covered by a copy aiming at covering another focus, the marginal part of users' view may be vague. To address this issue, we produce a copy covering these two near focuses to prevent this problem. Figure \ref{draw:focusmerging} shows focus merging technology.

\para{Prefetch Strategy:} The second optimization is prefetch strategy. One can construct a probabilistic model from user data and apply it to predict the movement of viewpoint. Based on its prediction of viewpoint, the server and the client will both leave some bandwidth to prefetch the predicted copies. If the users behave as predicted, they will immediately see the high-resolution part without any lag phase. %However, this optimization needs brand new application protocols to support. 

	\presec
\section{Experimental Evaluation}
\postsec
\label{Experiment}

\subsection{Experimental Setup}

\para{User-watching Traces:} We use real user-watching trace collected by \texttt{Kandao Technology Co., Ltd.} \cite{Kandao} to simulate the runtime bandwidth and flow switching behaviors.
To objectively show the performance of FISF and the state-of-the-art, we select five different kinds of VR videos to carry out experiments.
These five videos are available at the website \cite{v0}.
The first video is a grouping dancing, which has multiple static focuses. The second is a video of constructing a bridge, which has a moving focus. The third is a VR broadcast, which has only one static focus.
%long-term live video streaming with no focus.
The fourth is a travel advertisement, which has no obvious focus. The fifth is a solo dance, which has a static focus.
We have released our source codes at GitHub \cite{Github}.

\para{Computer Setting:} We run simulation experiments on a HP OMEN Notebook PC 15 with 8 CPU cores and 16 GB memory.

\vspace{0.05in}
\para{Metrics:} We define four metrics to evaluate the transmission performance.
%1) \texttt{Switching number:} defined as the number of switching among different copies.\\ 
%2) \texttt{Standstill Time:} defined as the lasting time of lag phases when the video is standstill.\\
%3) \texttt{High quality rate:} defined as $T_h/T_t$, where $T_h$ denotes the time during which the users watch high-resolution areas, and $T_t$ denotes the total time of the watching trace.\\
%4) $\mathbb{\alpha}$: $\alpha$ is defined as $n/N$, where $n$ denotes the number of bytes transmitted using different algorithm, and $N$ denotes the number of bytes when transmitting the $360^\circ\times180^\circ$ high-resolution videos.\\ 

\begin{itemize}[leftmargin=1\parindent]
\vspace{-0.09in}
\item \texttt{Switching number:} defined as the number of switching among copies.

\vspace{-0.09in}
\item \texttt{Standstill Time:} defined as the lasting time of lag phases when the video is standstill.

\vspace{-0.09in}
\item \texttt{High quality rate:} defined as $T_h/T_t$, where $T_h$ denotes the time during which the users watch high-resolution areas, and $T_t$ denotes the total time of the watching trace.

\vspace{-0.09in}
\item $\mathbb{\alpha}$: $\alpha$ is defined as $n/N$, where $n$ denotes the number of bytes transmitted using different algorithm, and $N$ denotes the number of bytes when transmitting the $360^\circ\times180^\circ$ high-resolution videos.
\end{itemize}

\vspace{-0.09in}
\presub
\subsection{Experimental Results}
\postsub

\begin{figure}[h]
\centering
\postfig
\includegraphics[width=\figwidthdraw]{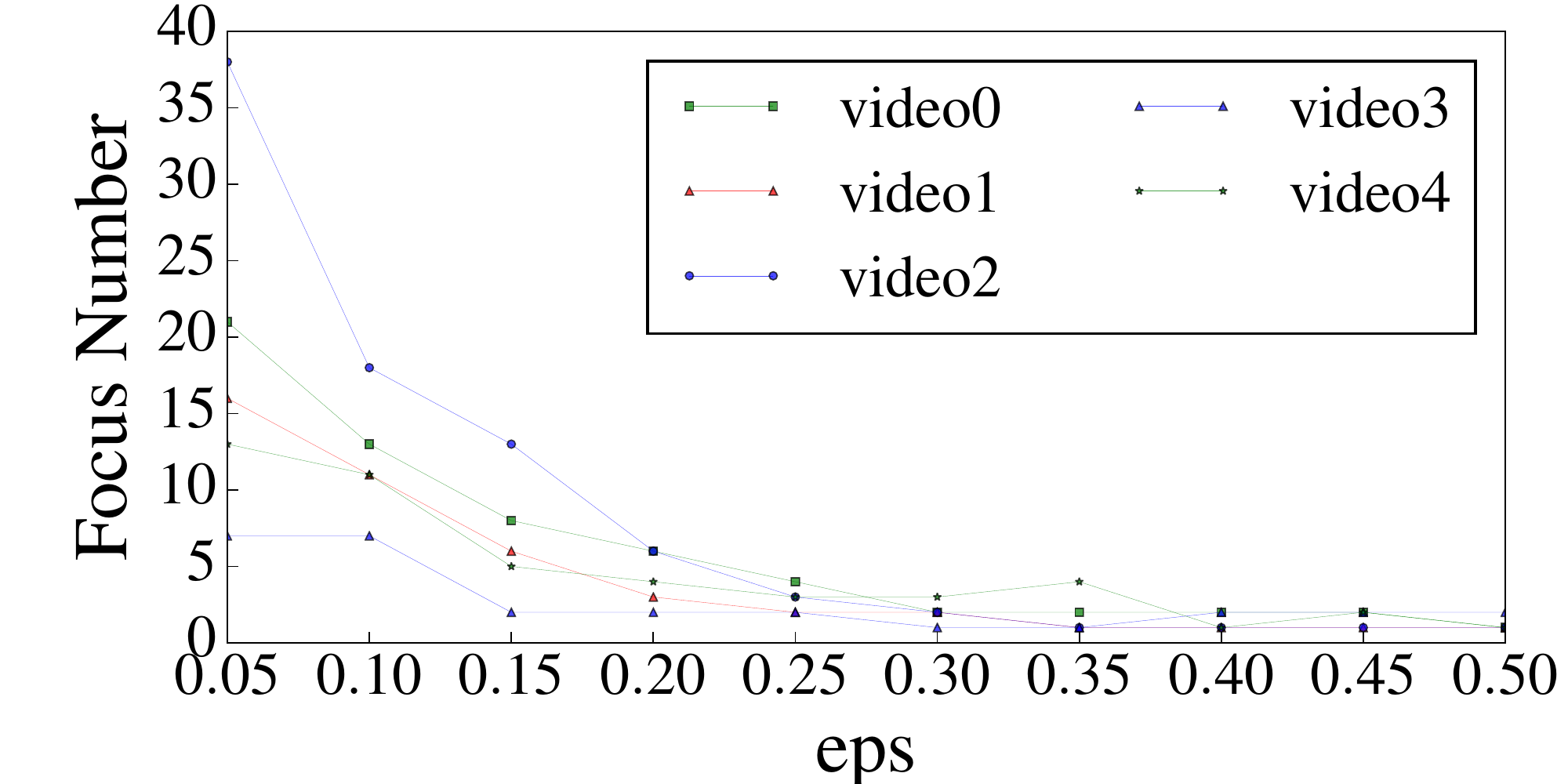}
\caption{Focus Number vs. eps.}
\label{eva:fnp}
\postfig
\end{figure}

\vspace{0.05in}
\para{Parameter selection:} As mentioned above, DBSCAN has two main parameters: \texttt{eps} and \texttt{min-samples}. The choice of these two parameters will significantly affect the accuracy of focus detection and the performance of FISF. Figure \ref{eva:fnp} shows the relation between focus number and eps. We can see that for each video, the focus number declines with the increase of eps. Our experiment also shows that the results do not have a clear relation to min-samples, thus we omit the corresponding experimental results. According to the experimental results, we preset eps as 0.3, min-samples as 100 for static FIST and eps as 0.2, min-samples as 30 for dynamic FIST.
%\textcolor{red}{Our experiment also show that the results are not sensitive to the parameter of ..., thus we omit the corresponding experimental results. From these two experiments, we can find the optimal parameters choice can be solved via ... but it is hard to applied in reality because of the expensive time cost. So we can combine the experiment result (as selection range) with the heuristic algorithm to decide the parameters shown in \ref{haha}.}

\begin{figure}[h]
\centering
\postfig
\includegraphics[width=\figwidthdraw]{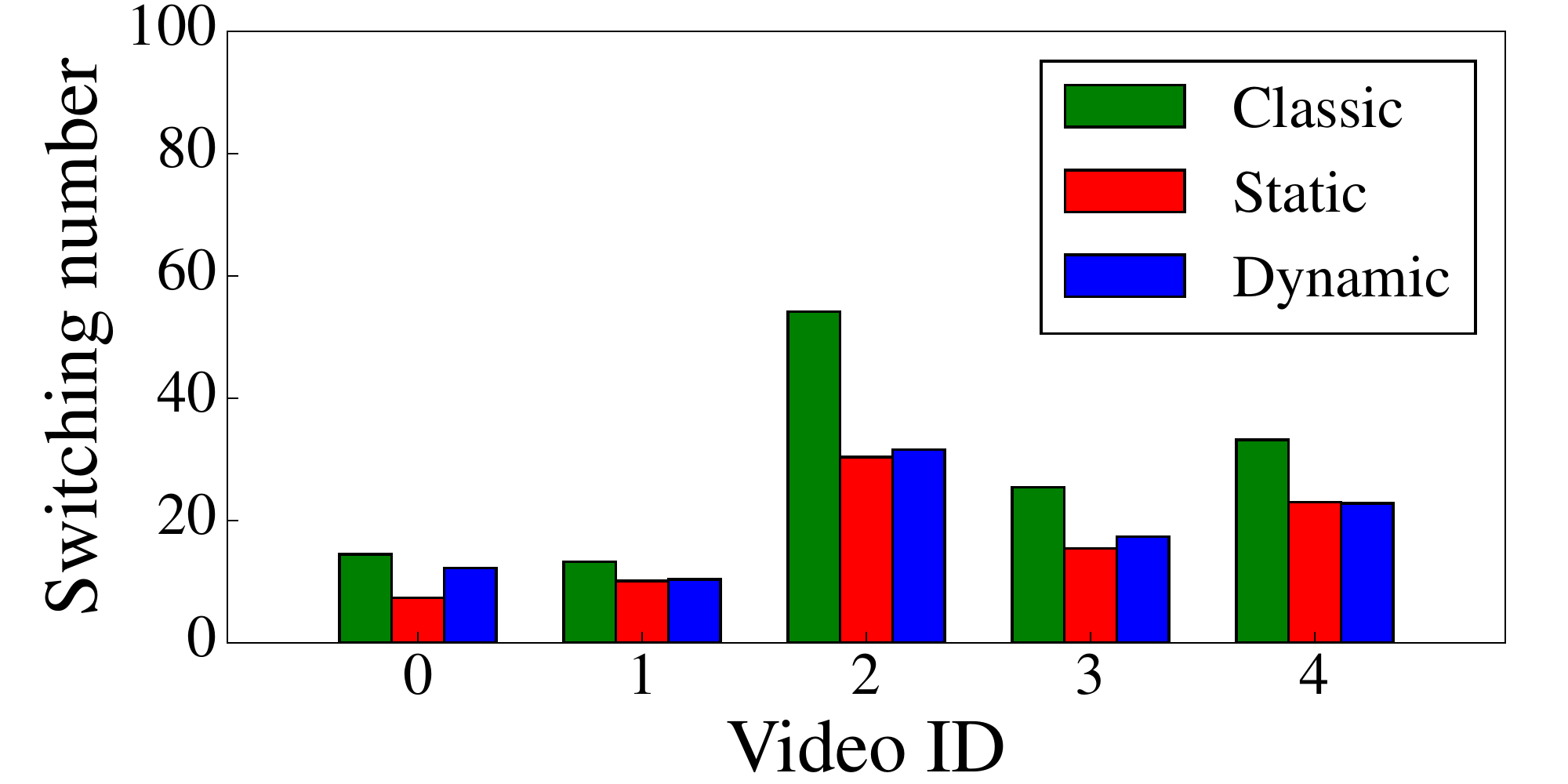}
\caption{User ID vs. Switching Number.}
\label{eva:fsn}
\postfig
\end{figure}

\para{Switching Number vs. Video ID :} 
Our results show that \textit{our static version can reduce the number of copy switching by $[23.4\%, 49.1\%]$, with a mean of $37.3\%$, and our dynamic version can reduce that by $[15.3\%, 41.7\%]$, with a mean of $28.3\%$}, compared to the classic algorithm.
As shown in Figure \ref{eva:fsn}, the x axis represents the video ID and y axis represents the number of copy switching. 

Copy switching may cause additional lag phases and computation cost. Thus the switching number should be reduced as much as possible.
For most videos, the number of focuses is usually small, and users tend to keep eyes around the focuses.
In this way, the number of copy switching will be reduced, and the standstill time will be reduced and computation resource will be saved.

\begin{figure}[h]
\centering
\postfig
\includegraphics[width=\figwidthdraw]{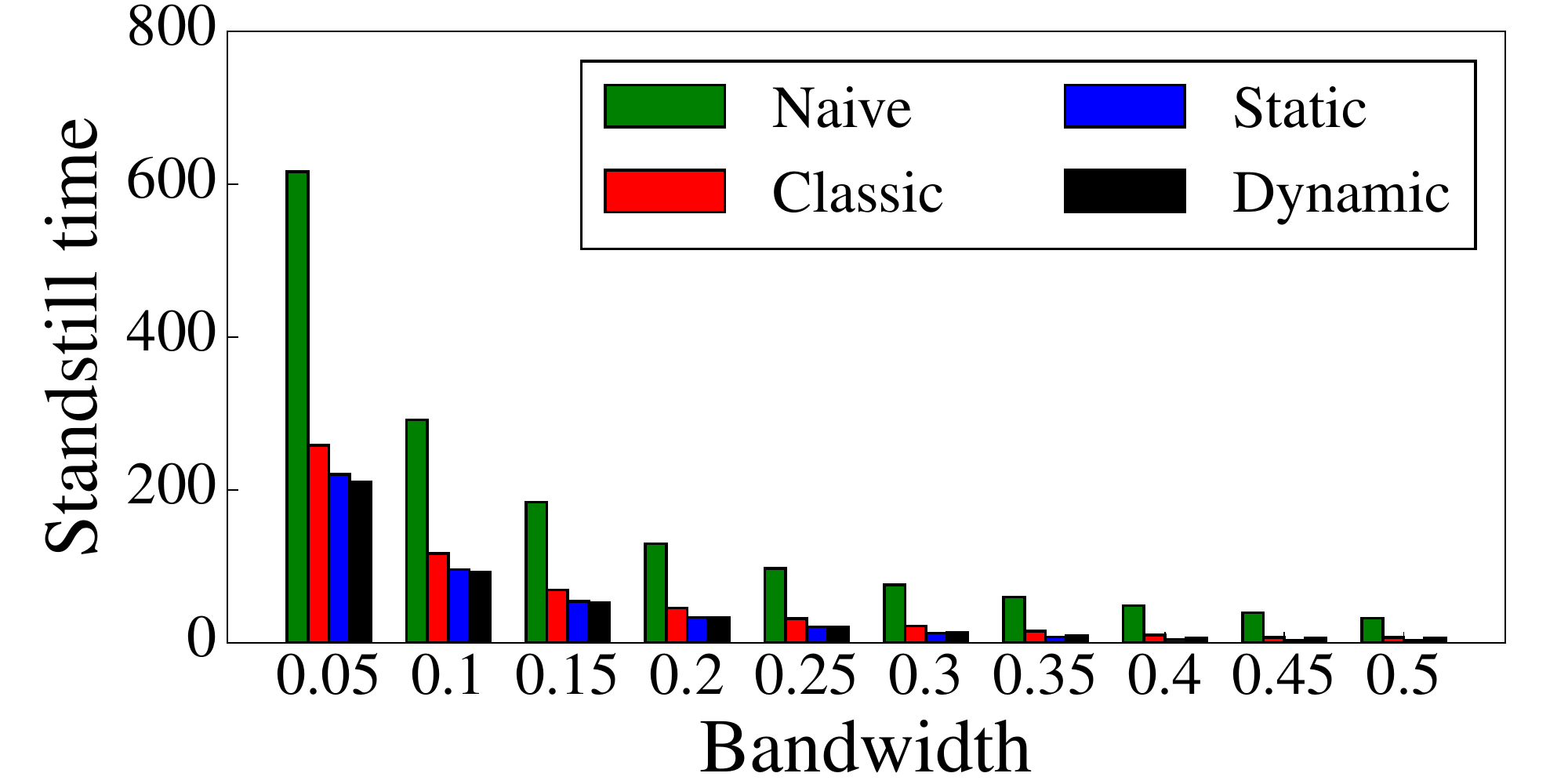}
\caption{Bandwidth vs. Standstill time.}
\label{eva:bandwidth2}
\postfig
\end{figure}

\vspace{0.05in}
\para{Standstill time vs. Video ID:} 
Our results show that \textit{our static version can shorten standstill time by $[14.9\%, 53.8\%]$, with a mean of $35.8\%$, and our dynamic version can shorten that by $[15.3\%, 40.9\%]$ , with a mean of $26.9\%$}, compared to the classic algorithm.
As shown in Figure \ref{eva:bandwidth2}, x axis represents the bandwidth limitation and y axis represents standstill time. Note that standstill time is relative, taking naive version as benchmark. The bandwidth limitation is also relative, and we suppose the bandwidth as $1.0$ with which users can watch the $360^\circ\times180^\circ$ high-resolution videos with no lag phases.

The bandwidth requirement of our algorithm is lower and the number of copy switching is smaller as well, thus our algorithm can significantly reduce standstill time and provide a more fluent watching experience for users.
%the standstill time of our static version is $14.9$\% to $53.8$\%, with a mean of $35.8\%$ shorter and that of dynamic version is from $15.3$\% to $40.9$\% , with a mean of $26.9\%$ shorter . Note that standstill time is relative, taking naive version as benchmark. The x axis is also relative and we suppose the bandwidth with which users can watch raw videos with no lag phase as $1.0$.

\begin{figure}[h]
\centering
\postfig
\includegraphics[width=\figwidthdraw]{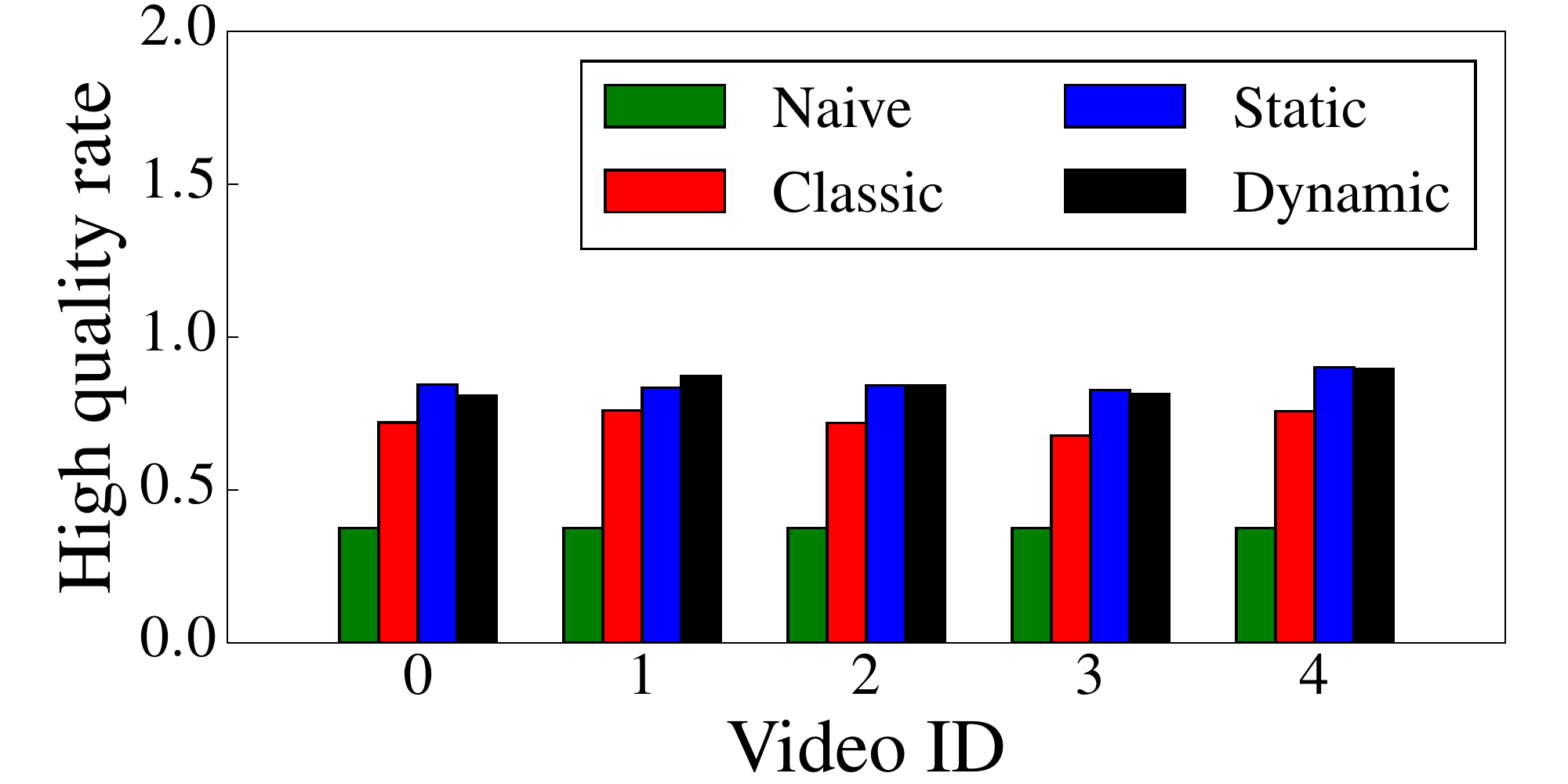}
\caption{User ID vs. High quality rate.}
\label{eva:ue}
\postfig
\end{figure}

\para{High Quality Rate vs. Video ID:} 
Our results show that \textit{our static version can improve high quality rate by $[9.7\%,  21.9\%]$ with a mean of $16.9\%$, and our dynamic version can improve that by $[12.1\%, 19.9\%]$ with a mean of $16.4\%$}, compared to the classic algorithm.
As shown in Figure \ref{eva:ue}, x axis represents the video ID and y axis represents the high quality rate. 
%High quality rate is one of the most important metrics to evaluate user experience. 
%Users will keep their eyes on the focuses in most of the watching time, and our method ensures that the area around the focuses can maintain high resolution. Thus our method can improve the high quality rate and user watching experience a lot.

%As shown in Figure \ref{eva:bandwidth1}, \textcolor{red}{x axis} representing the bandwidth and y axis representing the total bytes transmitted, the static version reduce the total data transmitted from $10.5$\% to $20.1$\%, with a mean of $15.1\%$, and the dynamic version reduces it from $14.6$\% to $20.4$\%, with a mean of $18.2\%$. Note that our data size is relative, taking that of naive version as benchmark.

\begin{figure}[h]
\centering
\postfig
\includegraphics[width=\figwidthdraw]{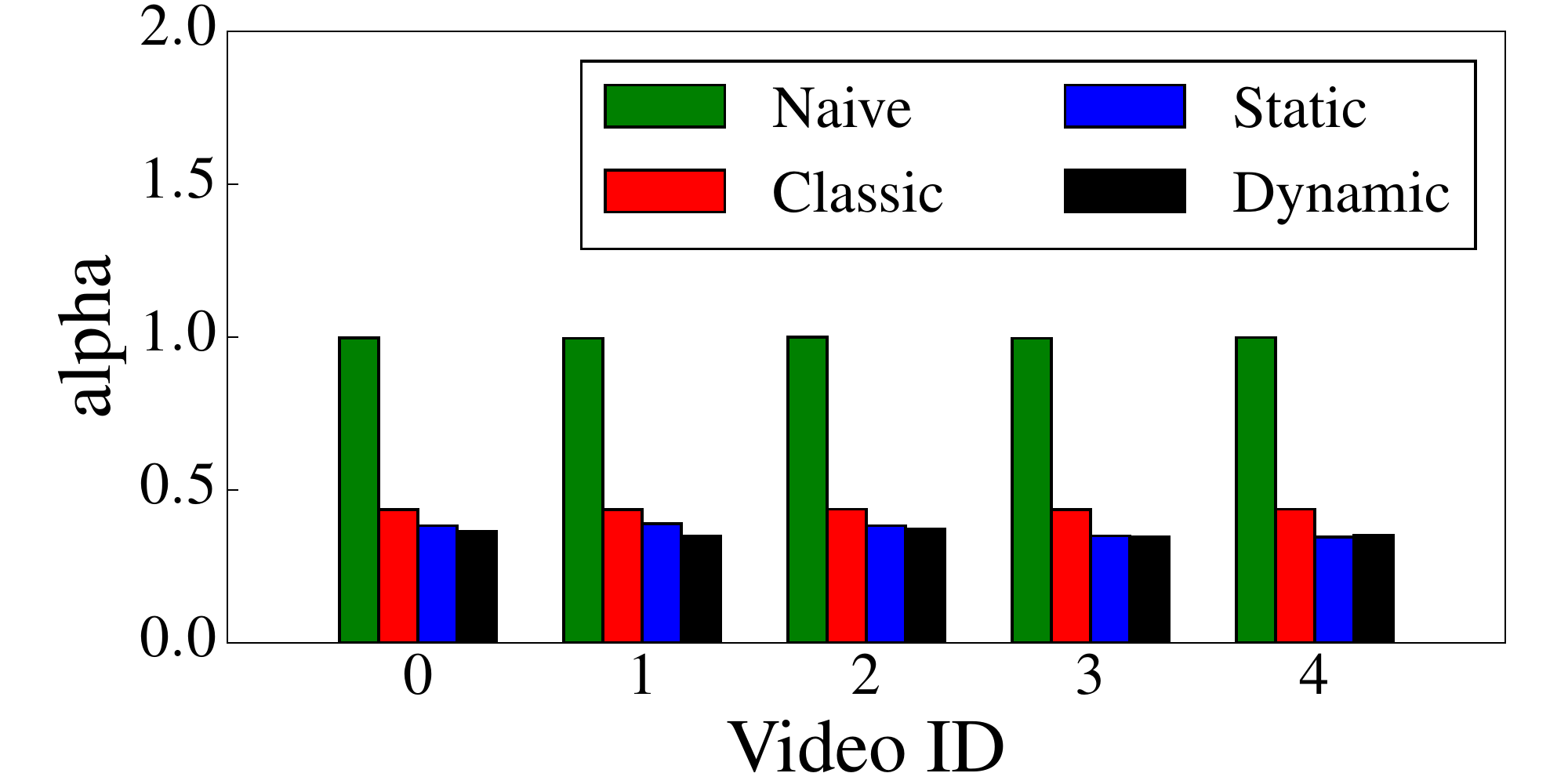}
\caption{Video ID vs. Ratio of transmitted bytes $\alpha$.}
\label{eva:bandwidth1}
\postfig
\end{figure}

\vspace{0.05in}
\para{$\alpha$ vs. Bandwidth:}
Our results show that \textit{our static version can reduce $\alpha$ by $[10.5\%, 20.1\%]$ with a mean of $15.1\%$, and our dynamic version can reduce that by $[14.6\%, 20.4\%]$ with a mean of $18.2\%$}, compared to the classic algorithm.
As shown in Figure \ref{eva:bandwidth1}, x axis represents the bandwidth and y axis represents $\alpha$. 
%$\alpha$ is defined as $n/N$, where $n$ is the number of bytes transmitted using different algorithm, and $N$ is the number of bytes when transmitting the $360^\circ\times180^\circ$ high-resolution videos.
%Our algorithm is based on the observation: when users keep their eyes on an area with a focus, they will be likely to ignore the other areas.
%Existing algorithm does not make use of the knowledge of video focus, thus can only produce video copies with a rectangle 90*120.
%As our algorithms are based on focuses, we can produce video copies with a circle 60*60
%
%
%Therefore, we can reduce the bandwidth requirement and
%
%Based on this observation, we can make high-resolution area smaller, and reduce the bandwidth requirement. It can significantly improve the watching experience of users with limited bandwidth.

	\vfill\eject
\presec
\section{Conclusion}
\postsec
\label{Conclusion}

Due to the requirement of low bandwidth and a superior user experience of panoramic VR videos, interactive streaming technology has drawn intensive attention in recent years. It has been put into practice by many corporations like Facebook \cite{Facebook}, Google, Microsoft, and DWANGO Co., Ltd.
 \cite{basicflowswitching}. However, the technology is far from mature because it brings about copy switching and degrades the user experience. Focus-based interactive streaming framework (FISF) points out a novel approach to addressing the problem by predicting behaviors of users, according to the characteristics of videos. It consists of a data-based video focus detection (VFD) , two versions of FIST, and two optimizations. Experimental results show that FISF significantly improves the user experience and reduces transmission bandwidth. 

To the best of our knowledge, \textit{this is the first time} that video focus detection based on real data analysis is used to optimize panoramic VR video transmission. There are still extensive future work about FIST, like parameter choice strategy and dynamic copy producing. We believe that FISF will be implemented and widely used to provide user-friendly and bandwidth-friendly transmission for panoramic VR videos in the near future.

\eject

	\balance
	%\vfill
	\bibliographystyle{abbrv}
	\bibliography{reference}
\end{document}